\documentclass[final,1p,times]{elsarticle} 

\usepackage{graphicx}
\usepackage{amssymb} 
\usepackage{amsthm} 
\usepackage{lineno}
\usepackage{amsmath} 
\usepackage{xspace}
\usepackage{hepnames}

\newcommand{\pt}{$p_{\rm T}$}

\newcommand{\GeVc}{\ensuremath{{\,\text{Ge\hspace{-.08em}V\hspace{-0.16em}/\hspace{-0.08em}}c}}\xspace}

\journal{Nuclear Physics A} 

\begin{document}

\begin{frontmatter} 

\title{b-jet Identification in PbPb Collisions with CMS}

\author{Matthew Nguyen (for the CMS Collaboration)}
\address{\'Ecole Polytechnique, Route de Saclay, 91128, Palaiseau, France}

\begin{abstract} 
The flavor dependence of jet quenching is a powerful handle to discriminate between models of parton energy loss in heavy-ion collisions. We demonstrate the capacity of CMS to identify jets initiated by bottom quarks using displaced vertices reconstructed in the silicon tracking system. The b-jet to inclusive jet ratio is measured in PbPb collisions and compared to pp collisions and simulations at the same center-of-mass energy. 
\end{abstract} 

\end{frontmatter} 


Identification of reconstructed jets from fragmentation of b quarks has so far not been performed in heavy-ion collisions.  However, recent CMS data demonstrate that non-prompt \PJgy mesons, i.e., those coming from decay of B hadrons, are indeed suppressed in PbPb collisions with respect to the pp expectation~\cite{Chatrchyan:2012np}.  This indirect measurement, which samples typical \PJgy \pt\ of values on the order of 10 \GeVc, provides a strong motivation to perform a more direct measurement using fully reconstructed jets.  Such a measurement would enable a direct comparison of b quark energy loss to that of inclusive jets at much larger values of jet \pt, where the flavor dependence of parton energy loss can be probed in detail.

CMS is described in detail in~\cite{cmsjinst}.  Jets formed from heavy flavor quark fragmentation can be tagged by the presence of displaced vertices, either by direct reconstruction of these vertices or by the impact parameter (i.e., the distance of closest approach to the primary vertex) of tracks originating from these vertices~\cite{CMS-PAS-BTV-11-004}.   Information from these tracks and vertices are typically combined into a quantity which optimizes their discrimination between heavy and light flavor jets.  In this analysis, we use a discriminator to tag b-jets which is based on the flight distance of the reconstructed secondary vertex (SV) with respect to the primary vertex of the interaction.  The efficiency of this SV tagging is then evaluated both directly from simulation, and with a data-driven method using a weakly correlated tagger based on the impact parameter of tracks associated to the jets.

The performance of lifetime-based tagging relies on the  high efficiency and low fake rate of reconstruction of charge particle tracks from displaced vertices. The standard heavy-ion tracking algorithm~\cite{hin10005} in CMS is largely restricted to the reconstruction of charged particles from the primary vertex.  To enhance the efficiency of tracks from secondary vertices, additional track reconstruction is performed. Secondary track reconstruction in central heavy-ion events is extremely resource intensive, due to the large number of possible hit combinations when incorporating hits over a sufficiently large region to find such displaced tracks.  To mitigate the number of possible track candidates, the additional track reconstruction uses reconstructed jets as seeds and limits the search window for tracker hits to a region defined around the jet axis.

Identification of b-jets is achieved using a discriminating variable, denoted Simple Secondary Vertex High Efficiency (SSVHE)~\cite{CMS-PAS-BTV-11-004}, which is based on the flight distance 
of reconstructed SVs.  Detailed studies of quantities related to SV reconstruction show that it is well modeled by PbPb simulations, lending confidence to the use of this discriminator to identify b-jets. Another discriminating variable, called Jet Probability (JP), is also used in this analysis.  The JP algorithm orders jet-associated tracks based on their impact parameter significance (impact parameter divided by its uncertainty) and calculates the likelihood that they come from the same primary vertex~\cite{CMS-PAS-BTV-11-004}.  The JP algorithm provides a measure of discrimination for nearly all b-jets, even in the case when no SV is reconstructed.   This property of the JP tagger is exploited to obtain a data-driven estimate of the SSVHE tagging efficiency~\cite{CMS-PAS-BTV-11-004}.  This is done by evaluating the fraction of jets from bottom fragmentation with and without the SSVHE tagging requirement, fitting the JP discriminator distribution with the same template fitting method described below.

\begin{figure}[!b]
\begin{center}
\resizebox{0.49\textwidth}{!}{\includegraphics{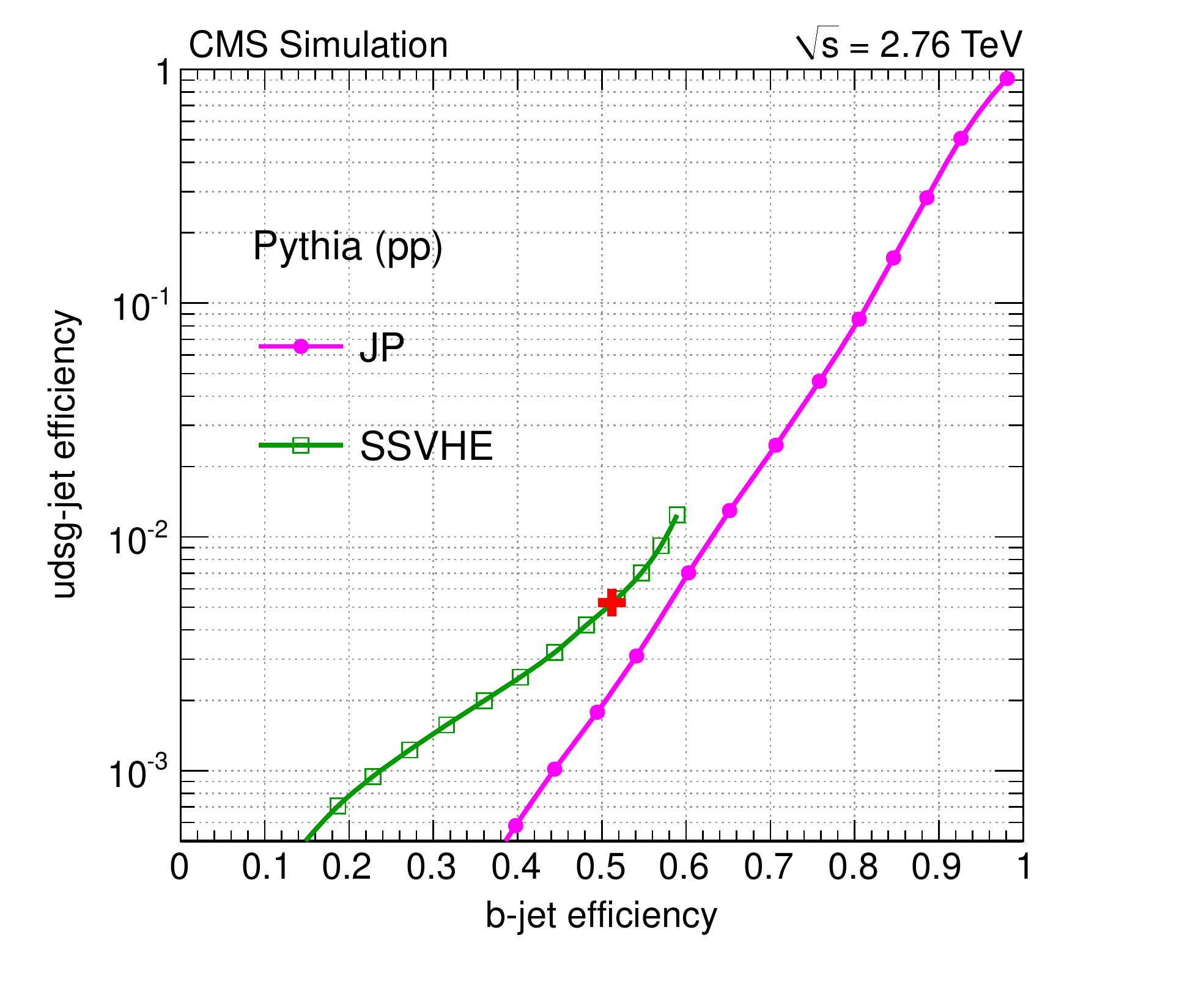}}
\resizebox{0.49\textwidth}{!}{\includegraphics{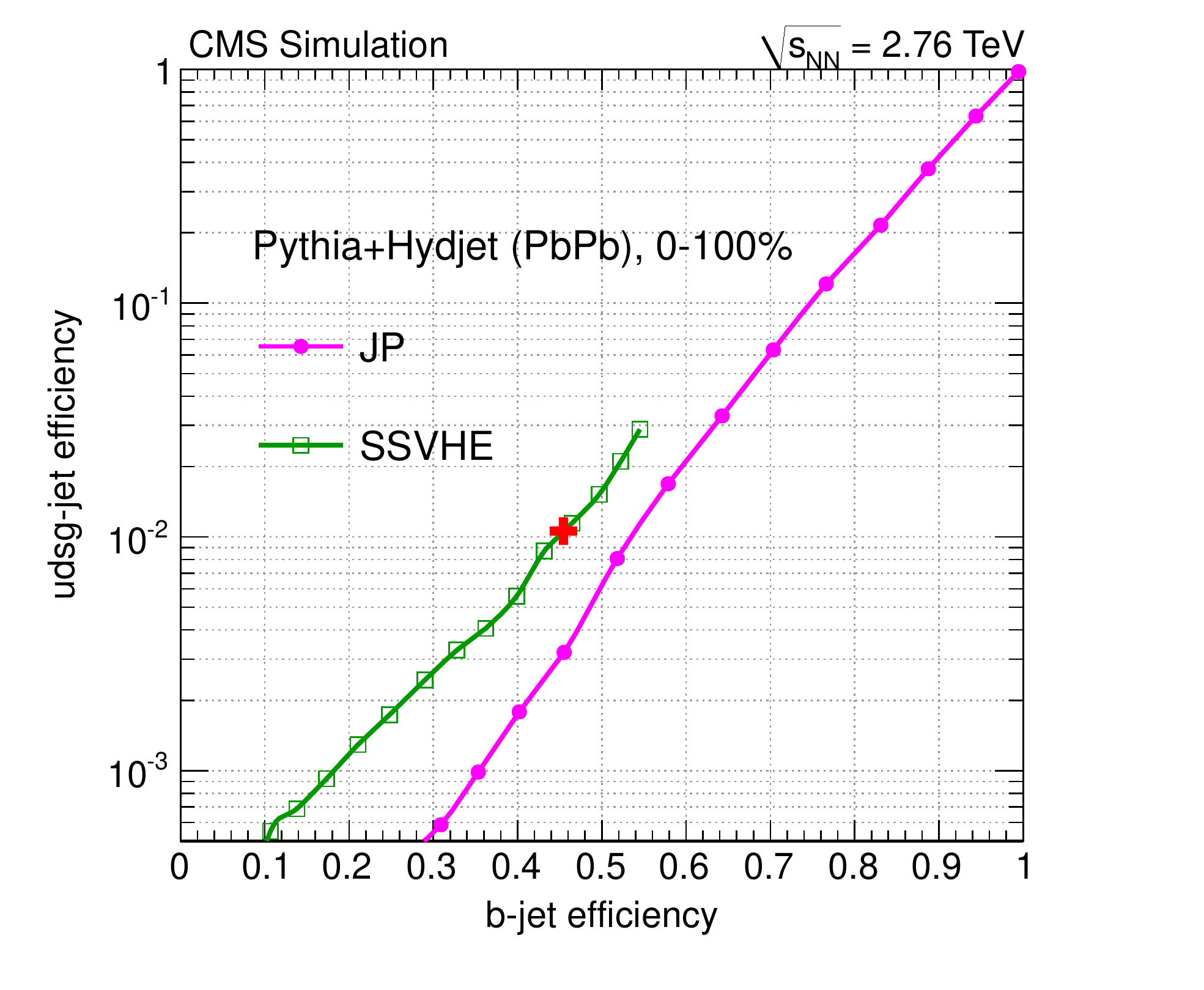}}
\end{center}
\caption[b-jet tagging efficiency vs. light jet mis-tag efficiency]{The b-jet tagging efficiency vs. the light jet mis-tag efficiency for simulated pp events from {\textsc{Pythia}} (left) and simulated PbPb events from {\textsc{Pythia}} embedded in {\textsc{Hydjet}} (right) for the SSVHE and JP discriminators.  The red cross marks the working point of the SSVHE discriminator used in this analysis.}
\label{fig:bVsL}
\end{figure}

The performance of the discriminators is typically benchmarked by plotting the b-jet tagging efficiency vs. the light (udsg) jet mis-tag efficiency, such that the resulting curves are independent of the underlying b-jet fraction.  Fig.~\ref{fig:bVsL} shows the b-jet tagging efficiency vs. the light jet mis-tag efficiency from {\textsc{Pythia}} and {\textsc{Pythia+Hydjet}}.  No centrality selection is applied to the {\textsc{Pythia+Hydjet}} sample, but its centrality distribution is weighted to match the distribution in data after the jet selection.  The two taggers used in this analysis, SSVHE and JP, are shown.  The performance of the b-tagging degrades somewhat with the increased multiplicity of PbPb collisions, giving roughly a factor of 3 poorer rejection of light jets for a b-jet efficiency of 50\%, relative to {\textsc{Pythia}} alone.  Despite the reduced performance, one is still able to achieve roughly a factor of 100 rejection of light jets for a b-jet efficiency around 50\%.  The charm rejection for this b-jet efficiency is about a factor of 10 (not shown).

\begin{figure}[!t]
\begin{center}
\resizebox{0.49\textwidth}{!}{\includegraphics{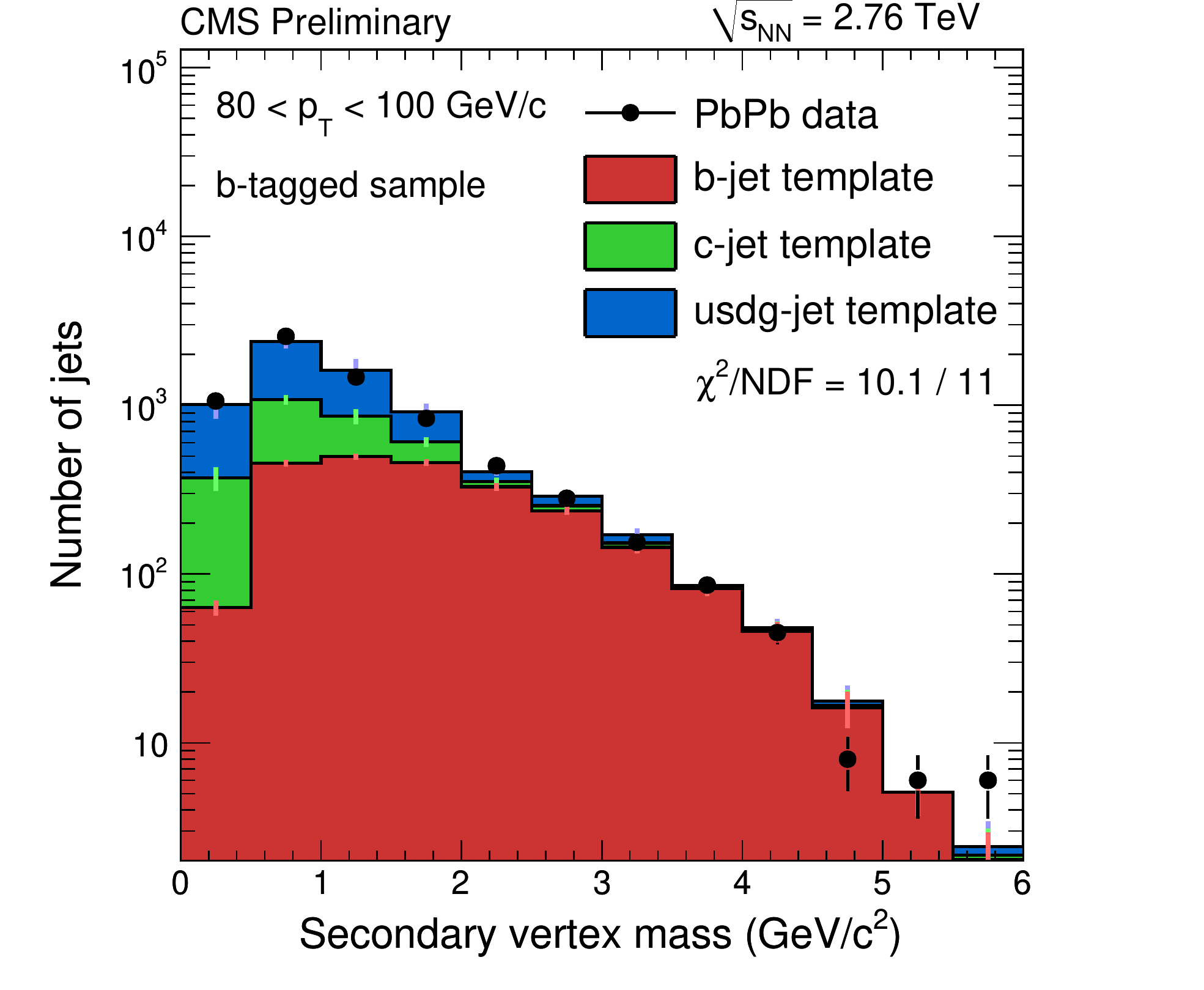}}
\resizebox{0.49\textwidth}{!}{\includegraphics{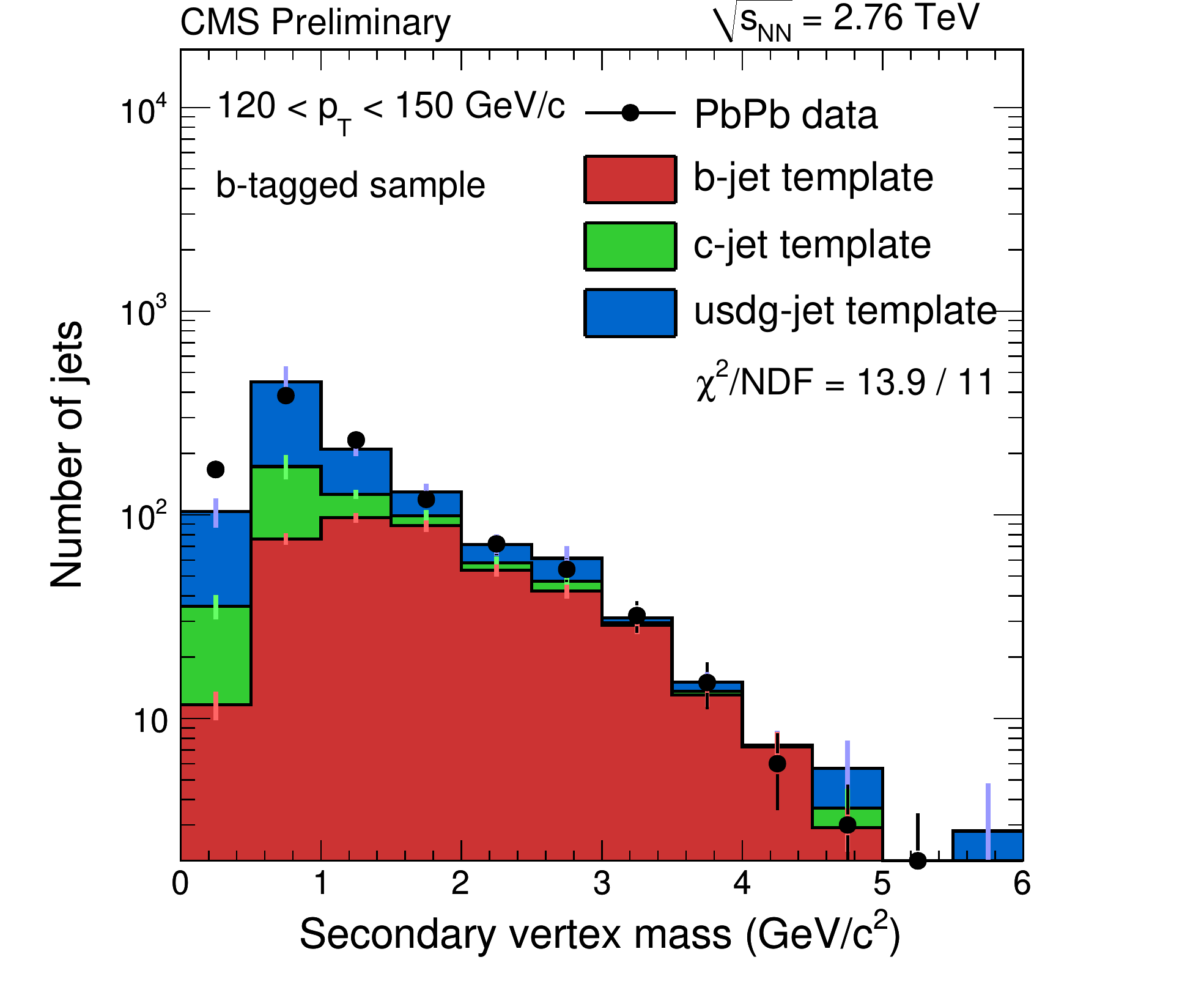}}
\end{center}
\caption{Template fits to the SV mass distributions in PbPb collisions, after tagging with the SSVHE discriminator for jets of 80 $<$ \pt\ $<$ \GeVc (left) and 120 $<$ \pt\ $<$ 150 \GeVc (right).}
\label{fig:fitPbPb_SVmass}
\end{figure}

The ratio of b-jet to inclusive jets (the ``b-jet fraction'') is extracted from the data using the SV mass.  The SV mass, which is simply the invariant mass of the charged tracks used in the SV reconstruction, provides 
good separation between the charm and bottom contributions.  Unbinned maximum likelihood fits are then performed where the b-jet and non-b-jet shapes are fixed by the simulation, but their normalizations are allowed to float.  Figure~\ref{fig:fitPbPb_SVmass} shows SV mass distributions in data and corresponding simulation templates for two of the jet \pt\ bins used in this analysis.  The ratio of the charm to light normalizations are fixed in the template fits according to their relative contributions in the input simulation.

\begin{figure}[!b]
\begin{center}
\resizebox{0.49\textwidth}{!}{\includegraphics{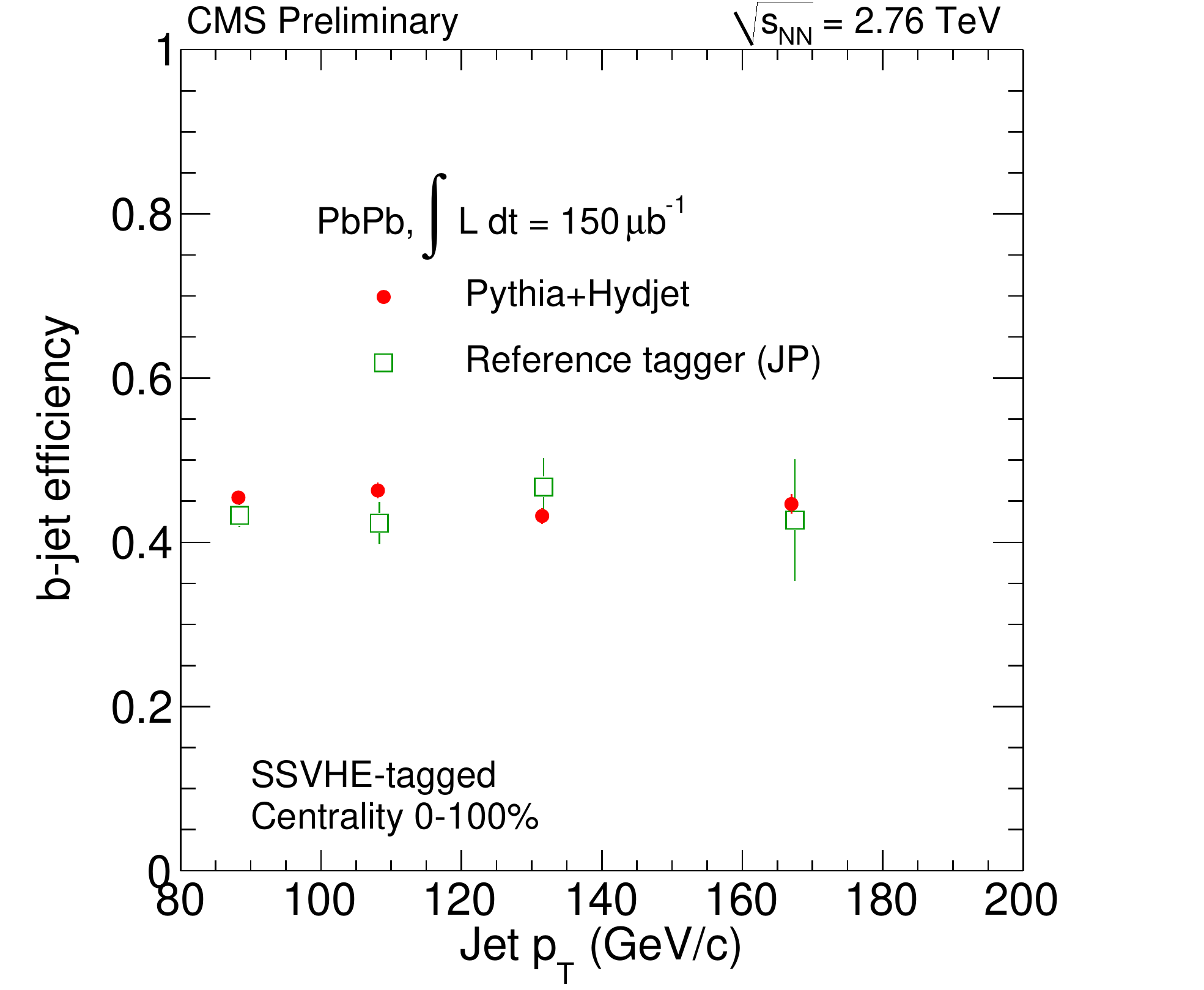}}
\resizebox{0.49\textwidth}{!}{\includegraphics{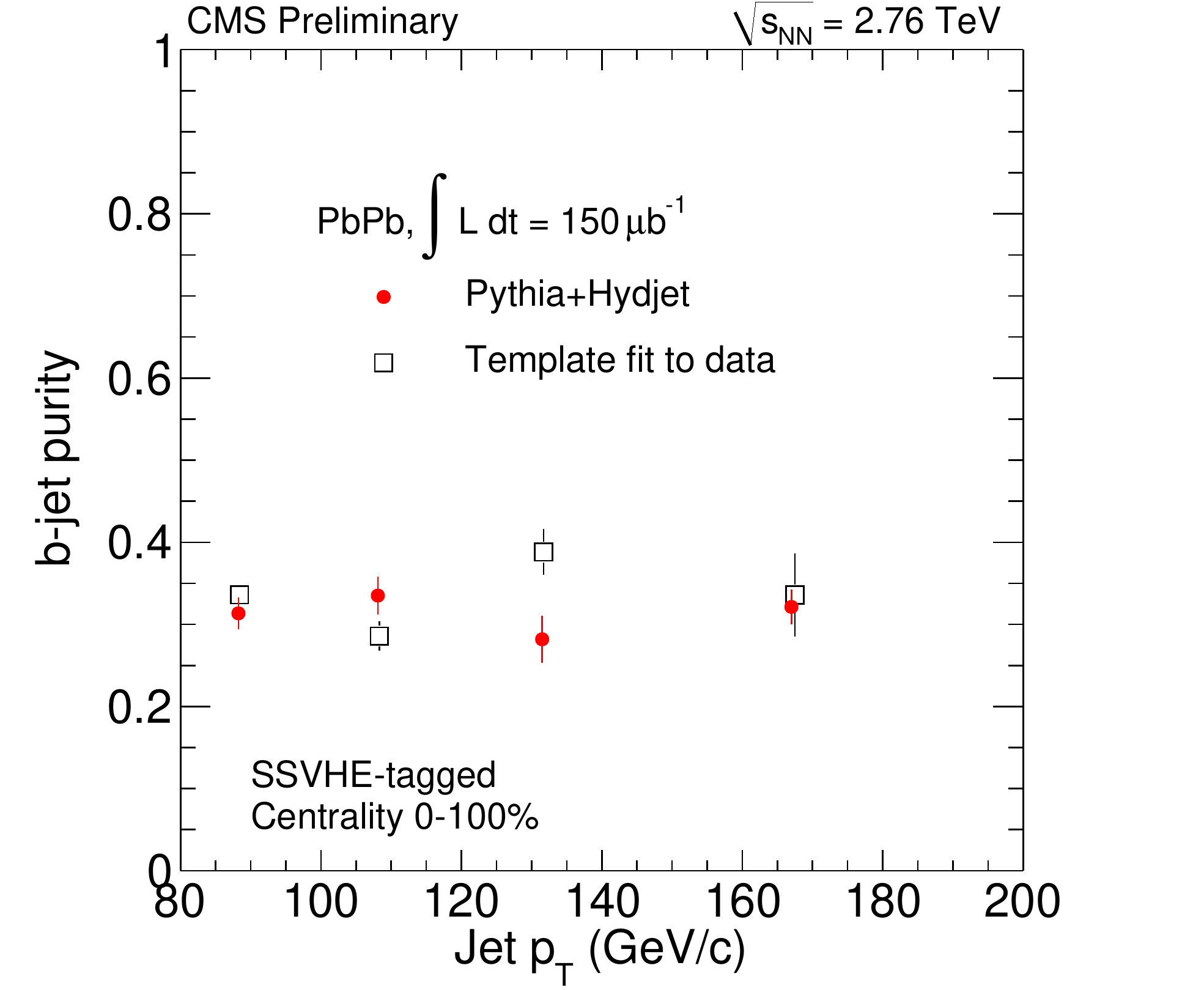}}
\end{center}
\caption[Efficiency and purity of secondary vertex tagging]{Left: The tagging efficiency of the SSVHE discriminator from simulation and from the reference tagger method.  Right: The b-jet purity extracted from template fits to the SV mass distributions, compared to the input simulation.}
\label{fig:ssvheEffPur}
\end{figure}

The SSVHE tagging efficiency is shown in the left panel of Fig.~\ref{fig:ssvheEffPur}.  The result obtained using JP as a reference tagger is compared to the result obtained directly from simulation.
The b-jet purity is the fraction of jets from bottom quarks in the SSVHE-tagged sample.
The right panel of Fig.~\ref{fig:ssvheEffPur} shows the b-jet purity of the tagged sample as a function of the jet \pt\ extracted from the template fits in Fig.~\ref{fig:fitPbPb_SVmass}.  These purity values are close to those from the input simulation. 

\begin{figure}[!t]
\begin{center}
\resizebox{0.49\textwidth}{!}{\includegraphics{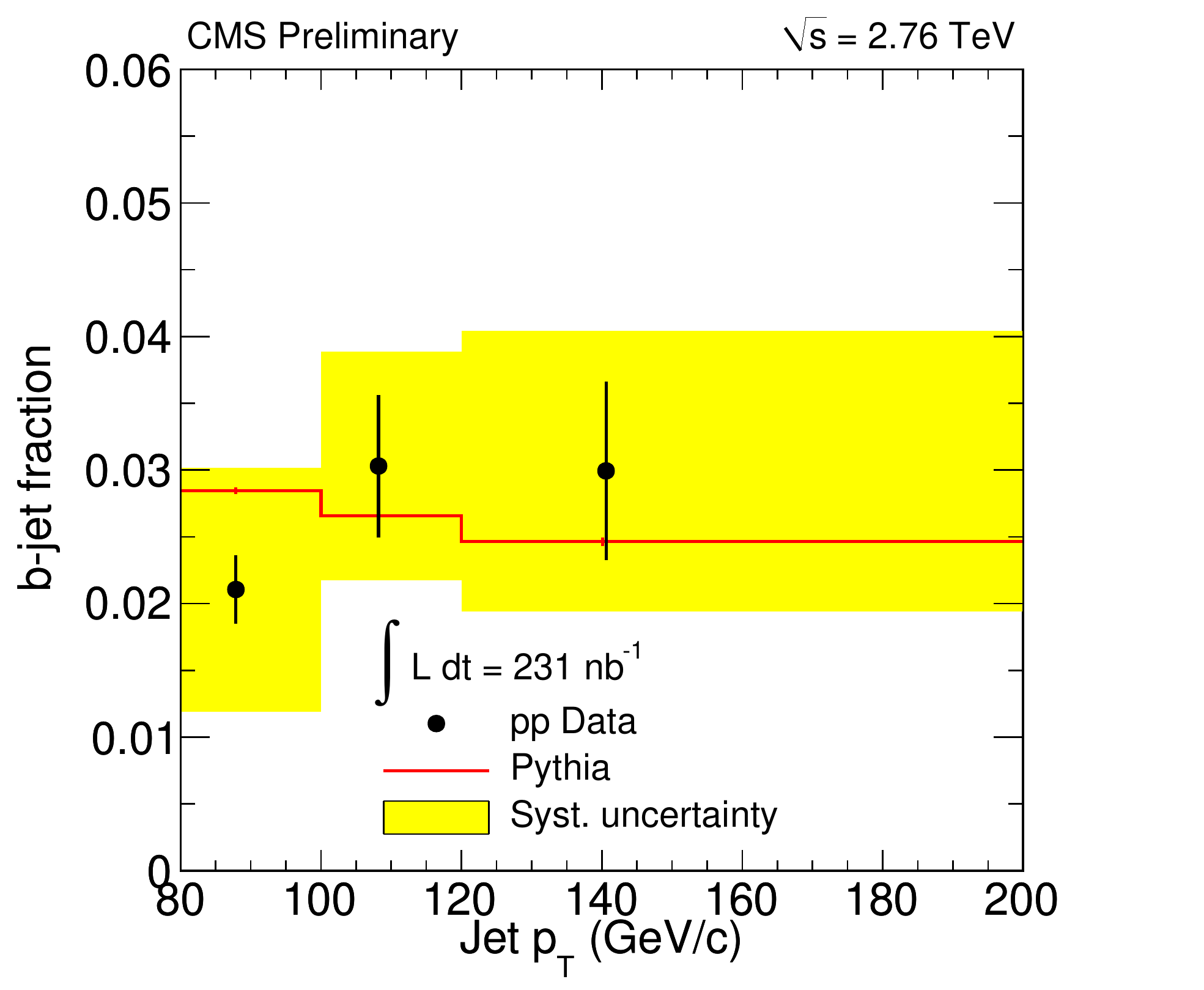}}
\resizebox{0.49\textwidth}{!}{\includegraphics{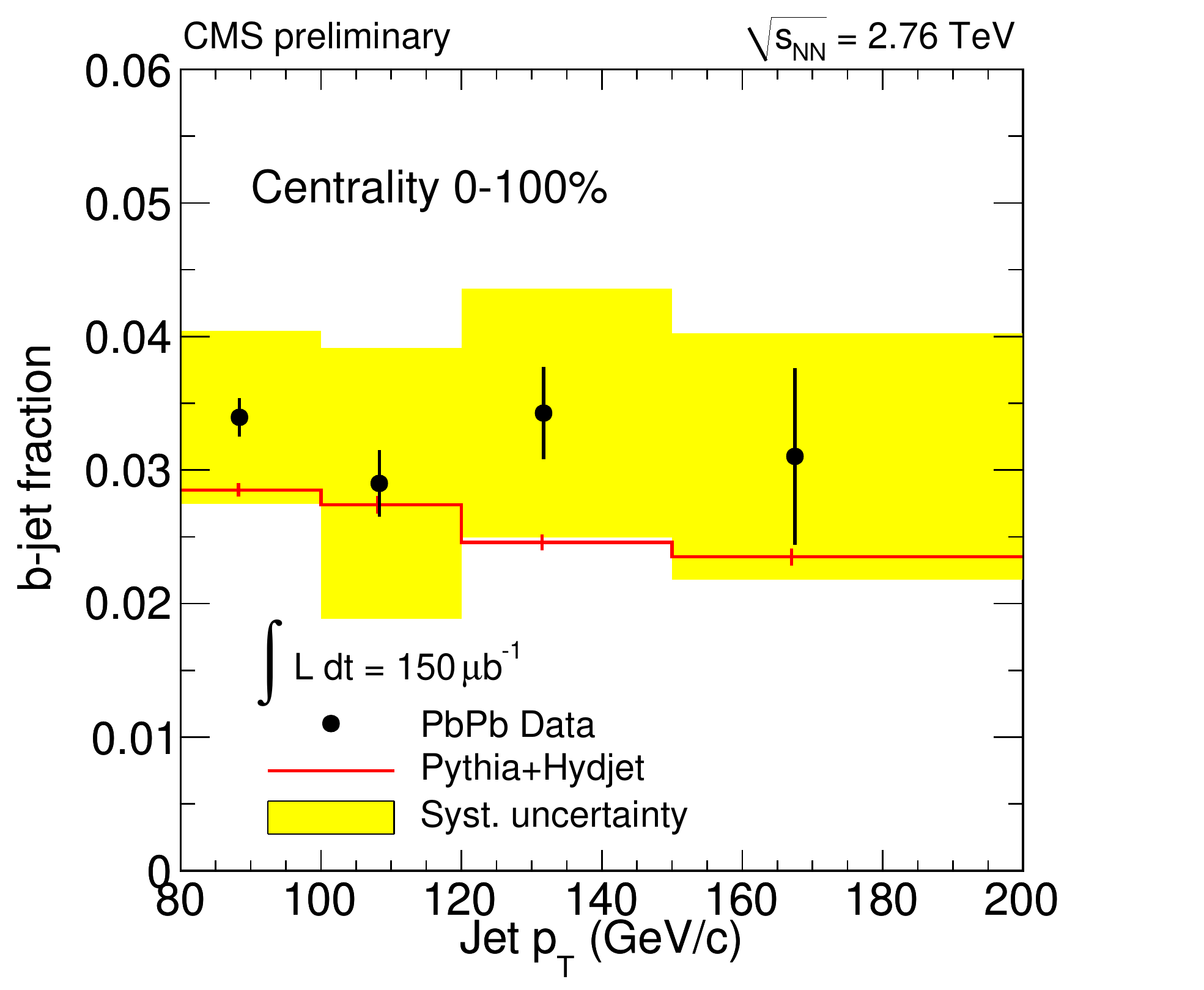}}
\end{center}
\caption[b-jet fraction in PbPb and pp vs jet \pt]{The b-jet to inclusive jet ratio in 0-100\% PbPb collisions (left) and pp collisions (right) as a function of jet \pt\ compared to {\textsc{Pythia}} embedded in {\textsc{Hydjet}} (PbPb) and {\textsc{Pythia}} (pp).}
\label{fig:bfraction}
\end{figure}

The b-jet to inclusive jet ratio is calculated from
$ (N^{\rm tagged}_{\rm jets} / N_{\rm jets}) (P/\epsilon)$,
where $N^{\rm tagged}_{\rm jets}$ and $N_{\rm jets}$ are the number of tagged jets and the total number of jets counted from the data, respectively, and $P$ and $\epsilon$ correspond to the purity and efficiency plotted in the left and right panels of Fig.~\ref{fig:ssvheEffPur}, respectively.
The left panel of Fig.~\ref{fig:bfraction} shows this ratio as function of jet \pt\ in 0-100\% PbPb centrality collisions.  The b-jet fraction is around 2.9--3.5\% with no significant \pt\ dependence, with an absolute uncertainty in the range of 0.6--1.1\%, increasing with jet \pt.  The values predicted by {\textsc{Pythia+Hydjet}}, which decrease from about 3\% to 2.5\% as a function of \pt, are consistent with the data.  The b-jet fraction in pp collisions as a function of jet \pt\ is shown in the right panel of Fig.~\ref{fig:bfraction}.  The values predicted by {\textsc{Pythia}} are consistent with the data.

There are several sources of systematic uncertainty which affect the b-jet fraction.  The uncertainty on the tagging efficiency is evaluated from the difference between the simulation and data-driven efficiencies.  A second category of uncertainty is composed of effects which influence the b-tagging purity.  One such uncertainty is the relative contribution from charm and light jets, which changes the shape of the non-b-jet template.  The relative normalization is fixed in our template fits, and then alternately allowed to float to obtain a systematic uncertainty.  
To determine the uncertainty on the modeling of the non-b-jet template shapes by the simulation, we compared to results obtained with a data-driven template formed by requiring an anti-b-tag on the JP discriminator.  An uncertainty on the stability of the template shapes was estimated by varying the working point of the SSVHE tagger. 
Other sources of systematic uncertainty considered are the b-jet and inclusive jet energy scales and the modeling of b-jet production in {\textsc{Pythia}}.

To within the fairly sizable uncertainties,  the data disfavor an extreme scenario in which b-jets suffer no energy loss in PbPb collisions.  A more detailed understanding of the jet \pt\ dependence of the b-jet to inclusive jet ratio and, more generally, the parton mass and flavor dependence of energy loss, would require a reduction of the statistical and systematic uncertainties.


\end{document}